\documentclass[prb,twocolumn,groupedaddress,showpacs,aps,floatfix,fleqn]{revtex4}

\usepackage{amsmath}
\usepackage{amstext}
\usepackage{epsfig}
\usepackage{graphicx}
\usepackage{dcolumn}
\usepackage{rotating}
\usepackage{draftcopy}
\usepackage{flafter}
\usepackage[below]{placeins}
\usepackage{ccaption}
\usepackage{soul}
\usepackage{wrapfig}
\usepackage{sidecap}
\usepackage{float}
\usepackage{subfig}
\usepackage{booktabs}
\usepackage{tabularx}
\graphicspath{{PP_figs/}}
\begin{document}
\bibliographystyle{apsrev}
\title{Predicted band structures of III-V semiconductors in wurtzite phase}
\author{A. De}
\author{Craig E. Pryor}
\affiliation{Department of Physics and Astronomy and Optical Science
and Technology Center, \\ University of Iowa, Iowa City, Iowa 52242}
\date{\today}
\begin{abstract}
While non-nitride III-V semiconductors typically have a zincblende structure, they may also form wurtzite crystals under pressure or when grown as nanowhiskers.
This makes electronic structure calculation difficult since the band structures of   wurtzite III-V semiconductors are poorly characterized.
We have calculated the electronic band structure for nine III-V semiconductors in the wurtzite phase using transferable empirical pseudopotentials including spin-orbit coupling.
We find that all the materials have direct gaps.
Our results differ significantly from earlier {\it ab initio} calculations, and where experimental results are available (InP, InAs and GaAs) our calculated band gaps are in good agreement.
We tabulate energies, effective masses, and  linear and cubic Dresselhaus zero-field spin-splitting coefficients for the zone-center states.
The large zero-field spin-splitting coefficients we find may lead to new functionalities for designing devices that manipulate spin degrees of freedom.
\end{abstract}
\maketitle
\section{Introduction}
Semiconductor nanowhiskers (NW) have attracted a tremendous amount of interest in recent years \cite{Haraguchi1992APL,Hiruma1993JAP,Ohlsson2001,Gutierrez2001,Poole2003APL,Regolin2007,Jabeen2009,Colombo2008}.
Much of it is due to their potential application in areas such as photovoltaic cells\cite{Hu2007,Tsakalakos2007,Czaban2009}, nano-electromechanical resonator arrays \cite{Henry2007}, microwave and THz detection \cite{Balocco2005,Gustavsson2008}, single photon detection\cite{Rosfjord2006,Zinoni2007,Dauler2007}, field-effect\cite{Cui2003}\cite{Greytak2004} and single-electron transistors\cite{Thelander2003}, and various other electronic and optoelectronic devices \cite{Bjork2002,Bjork2002b,Panev2003,Duan2001,Samuelson2004}.
NWs are also interesting because 1D systems can be made using material combinations for which large lattice mismatches prohibit quantum well structures, allowing greater freedom in material combinations for device engineering.

In contrast to bulk non-nitride III-V materials which are usually
zincblende (ZB), NWs predominantly crystalize in the wurtzite (WZ) phase
\cite{Koguchi1992JJAP,Mattila2006,Tomioka2007JJAP}. Several
different theoretical explanations for this behavior have been
proposed. Recent calculations suggest that the WZ phase is
energetically favorable for small NW radii
\cite{Akiyama2006PRB,Galicka2008JP}, although this does not account
for all NW radii for which the WZ phase is experimentally observed.
Calculations based on an empirical nucleation model indicate
that WZ formation is favored for certain ranges of the interface
energies\cite{Glas2007}. {\it Ab initio} calculations indicate that
the WZ phase is favored due to the accumulation of electrons at the
interstitial site containing the Au catalyst \cite{Haneda2008},
while other calculations  show that the polytype is determined by
growth kinetics\cite{Dubrovskii2008PRB}. It should be noted that
these different mechanisms are not necessarily mutually exclusive.

Theoretical understanding of the electronic and optical properties of semiconductor nanostructures is based  on a knowledge of the electronic properties of bulk materials. However little is known about the electronic band structure of III-V semiconductors in WZ phase since most do not naturally occur as bulk crystals.
Moreover, the NWs often contain sections of ZB material, forming heterostructures out of the differing band structures of the two polytypes\cite{Hiruma1993JAP,Xiong2006NL,Bao2008NL,Soshnikov2005}.
Band structures of WZ III-V semiconductors have been calculated using density functional theory (DFT) in the local density approximation (LDA)\cite{Murayama1994, Yeh1994}.
Since the LDA underestimates band gaps the WZ band structure can not be directly determined, and instead calculations of WZ and ZB are typically compared to obtain the differences between the two polytypes.
The band structures of GaAs and InAs in WZ phase have also been calculated using the GW method\cite{Zanolli2007},  giving somewhat different results than those from the LDA.
In addition to the inherent errors in {\it ab initio} band gaps,  all of the above calculations neglected the spin-orbit coupling, which is known to significantly alter the valence band structure of semiconductors.

In this paper we present calculations of the bulk electronic band structures of the nine non-nitride III-V semiconductors in WZ phase using empirical pseudopotentials including spin-orbit coupling.
These calculations are based on transferable model pseudopotentials assuming ideal WZ structure.
The spherically symmetric ionic model potentials are first obtained by fitting the calculated bulk ZB energies to experimental energies at high symmetry points.
The band structure of the WZ polytype is then obtained by transferring the model pseudopotentials to the WZ pseudopotential Hamiltonian  using the appropriate crystal structure factors.

This method has been proven to be very successful in obtaining the bulk band structures of semiconductor polytypes\cite{Bergstresser1967,Joannopoulos1973,Foley1986,Bandic1995,Pugh1999,Pennington2001,Fritsch2003,Fritsch2006,Cohen.book}.
The anion and cation pseudopotentials are specific to each material and are only transfered between polytypes.
Therefore the model potentials should be transferable between ZB and WZ polytypes due to the similarities in their crystal structures.
In both structures all of the nearest neighbors and nine out of the twelve second nearest neighbors are at identical crystallographic locations \cite{Birman1959} while the the second nearest neighbors are equidistant.

This paper is organized as follows.
In section \ref{sec:WZvZB} we outline the similarities and differences between ZB and WZ crystal structures as well as the direct correspondence between  high symmetry $k$-points in the two polytypes.
In section \ref{sec:PP-method} we describe the transferable pseudopotential method.
In section \ref{sec:results} we present the calculated band structures, their respective density of states (DOS), and effective masses.
Finally, we summarize the results in section \ref{sec:summary}.

\section{Wurtzite vs Zincblende}\label{sec:WZvZB}
\subsection{Crystal Structure} \label{sec:crystalStructure}
\begin{figure}
  \subfloat[]{\label{fig:ZB_view111}\includegraphics[width=0.5\columnwidth]{ZB_111.pdf}}
  \subfloat[]{\label{fig:WZ_view111}\includegraphics[width=0.5\columnwidth]{WZ_111.pdf}}
  \\
  \subfloat[]{\label{fig:ZB_bz}\includegraphics[width=0.5\columnwidth]{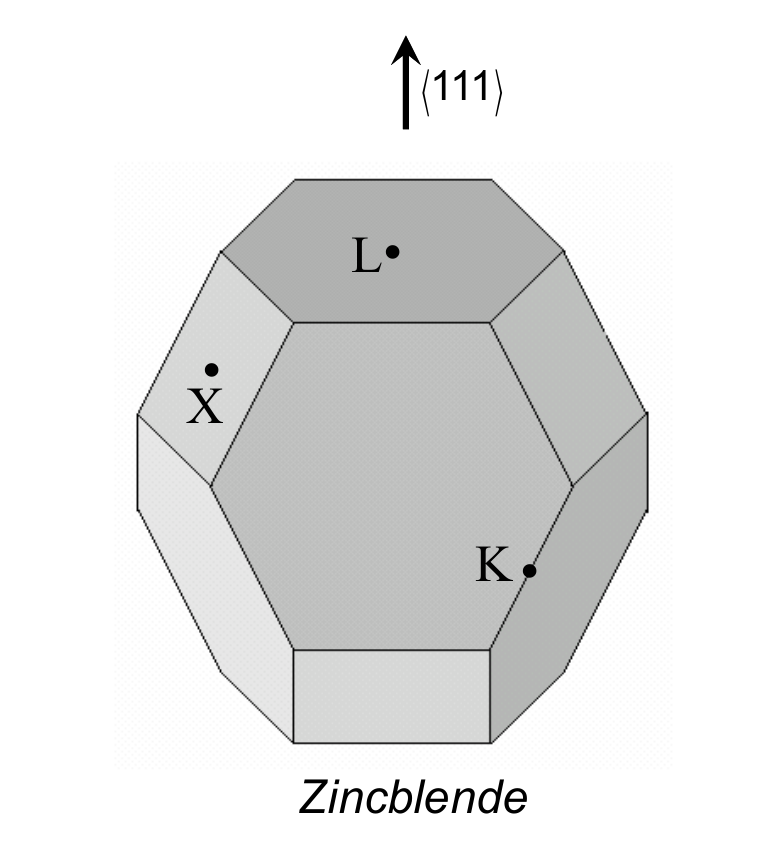}}
  \subfloat[]{\label{fig:WZ_bz}\includegraphics[width=0.5\columnwidth]{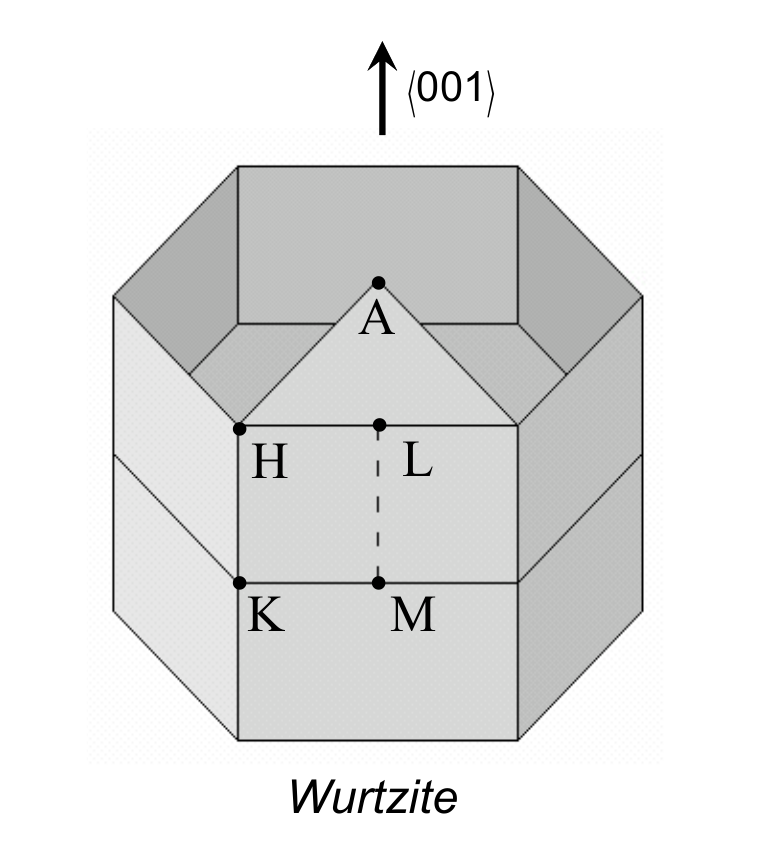}}
\caption{{\bf(a)} Staggered configurations of atom-1, as viewed
along [111] for ZB {\bf (b)} Eclipsed configuration of atom-1 as
viewed along [111] for WZ. Note that 9 out of the 12 second nearest
neighbors are in the same position. The other 3 are rotated by
$\pi$/3. {\bf (c)} Brillouin zone for ZB. {\bf (d)} Brillouin zone
for WZ. $\Gamma$ is at the center of the Brillouin zones.}
\label{fig:ZBvsWZ}
\end{figure}

The ZB crystal is formed by two interpenetrating face centered cubic
(FCC) Bravais lattices (each of a different atomic species), whereas
the WZ structure is constructed from two interpenetrating
hexagonal close-packed (HCP) lattices. The differences between the
two structures are best understood by viewing along the [111] direction
(Fig. \ref{fig:ZBvsWZ}a,b), along which both look like stacked hexagonal layers.
The atoms are identical within each layer, and the layers alternate between the anion and the cation.
For the ideal WZ crystal the lattice constant is given by $a_{WZ}=a_{ZB}/ \sqrt{2}$
and the lattice constant along the $c$-axis (axis perpendicular to the hexagon) is related to the in-plane lattice constant by $c=(8/3)~a_{WZ}$.
Since the WZ crystal is  tetrahedral, the nearest neighbors are the same in the two polytypes.
In addition, we see from Fig. \ref{fig:ZBvsWZ}  that nine of the twelve nearest neighbors in WZ are the same as in ZB.
These structural similarities suggest that the local electronic environment will be the same in the two crystals, and therefore the crystal potentials will be nearly identical in WZ and ZB.
In WZ the type-1 atoms are located at $(0,0,0)$ and $\frac{2}{3} {\bf a_1}+\frac{1}{3} {\bf a_2} + \frac{1}{2} {\bf a_3}$, while the type-2 atoms are located at $u {\bf a_3}$ and $\frac{2}{3}{\bf a_1} + \frac{1}{3} {\bf a_2} + (\frac{1}{2}+u){\bf a_3}$, where the primitive lattice vectors are ${\bf a}_1=(1,\sqrt{3},0) a_{WZ} / 2$, ${\bf a}_2=(1,-\sqrt{3},0) a_{WZ}/2$ and ${\bf a}_3=(0,0,c)$.
The parameter $u=3/8$ for the ideal WZ crystal structure, which we will assume throughout this paper.
This assumption is supported by the fact that experiments show that WZ GaAs is very close to ideal\cite{McMahon2005}.

\subsection{Band Structure}\label{sec:bandstructure}
\begin{figure}
\centering
\includegraphics[width=0.95\columnwidth]{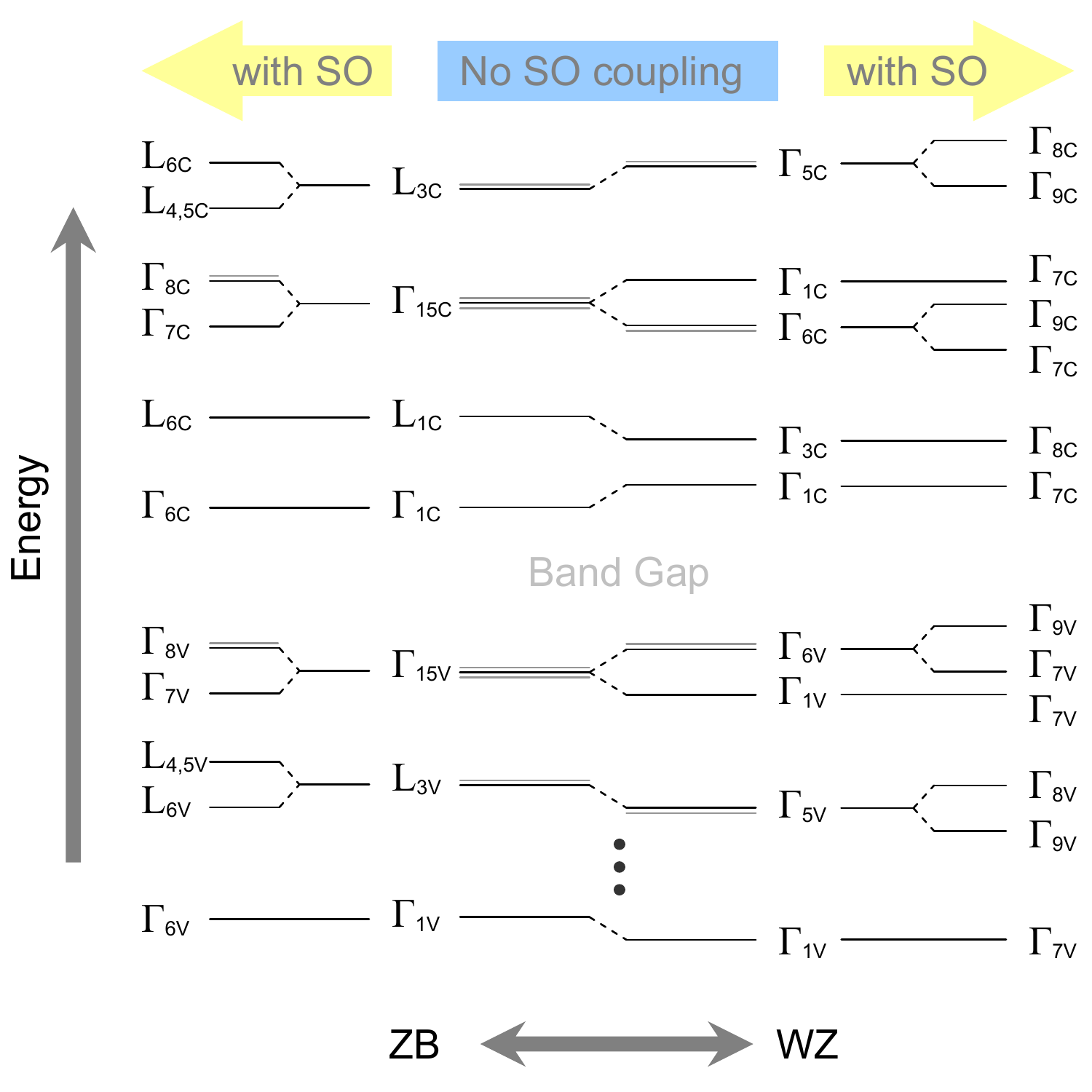}ind
\caption{ Schematic showing the correspondence between
energy levels at the $L$ and $\Gamma$ points in ZB and the $\Gamma$
point in WZ with and without spin-orbit coupling \cite{Murayama1994}. The
dashed lines show the correspondence between the states. Additional
degenerate levels are shown in grey. } \label{fig:levels}
\end{figure}

Due to the similarities of the two crystals, many of  the high symmetry points in the Brillouin zones of  ZB and WZ are related to each other, and an understanding of their correspondences is  useful for understanding trends in the band structures of the two polytypes.
Fig. \ref{fig:levels} shows the relationships among the zone
center states in WZ and the corresponding $L$ and $\Gamma$ points in ZB, both with and without spin-orbit coupling.
One of the most important features is that in the empty lattice approximation the L-point in ZB is zone-folded to the $\Gamma$-point in WZ.
As a result, in the absence of spin-orbit coupling, the $\Gamma_1$, $L_1$ and $L_3$ states in ZB correspond to $\Gamma_1$, $\Gamma_3$ and $\Gamma_5$ respectively in WZ (Fig. \ref{fig:levels}).
Because of this folding over of the $L$-valley, indirect gap ZB materials with an $L$ valley conduction band minimum would be expected to have a direct gap in the ZB phase unless the energy of the state was significantly shifted by the crystal potential.

The states at the top of the valence band in WZ also have some important differences with their ZB counterparts.
In the absence of spin-orbit coupling, the hexagonal crystal field of WZ splits the $p$-like $\Gamma_{15}$ state of ZB into a four-fold degenerate  $\Gamma_6 $ and a doubly degenerate $\Gamma_1$.
In terms of the $p$-orbitals, these states are $p_z\rightarrow\Gamma_1$ are $p_x,p_y\rightarrow\Gamma_6$.
With the inclusion of spin-orbit coupling, $\Gamma_{6v}$ splits into the $\Gamma_{9v}$ heavy-hole and the $\Gamma_{7v}$ light-hole.
Therefore, all zone center states in WZ belong to either $\Gamma_7$, $\Gamma_8$, or $\Gamma_9$.
There are similar correspondences between the high symmetry directions of the two crystals.
The symmetry line  $\Lambda ( \Gamma\rightarrow L$) in ZB corresponds to the $\Delta(\Gamma \rightarrow A)$ line in WZ\cite{Birman1959}.

\subsection{Spin Splitting}

\begin{figure}[ht]
\centering
\includegraphics[width=0.7\columnwidth]{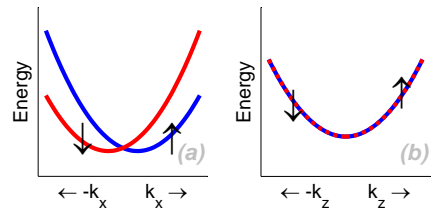}
\caption{({\bf a}) Zero field Dresselhaus spin splitting in WZ along
$k_x$ for a $\Gamma_7$ conduction band. ({\bf b}) The zero field
spin splitting is not seen along the $k_z$ direction. }
\label{fig:WZ_spin}
\end{figure}

For certain individual states in crystals lacking inversion symmetry, spin-orbit coupling causes a splitting of spin-up and spin-down states which leads to $E_{\uparrow}(-{\bf k})\neq E_{\uparrow}({\bf k})$ ( see Fig. \ref{fig:WZ_spin}). The states at $\mathbf k = 0 $ are still two-fold degenerate, resulting in a nonvanishing $\nabla_k E({\bf k})$ at the origin for certain crystallographic directions. In the case of WZ, the states remain spin-degenerate for $k$  along the $c$-axis
because at any point along the $k_z$ direction ($\Gamma\rightarrow A$), the crystallographic point group is $C_{6v}$\cite{Casella1959}. As all the irreducible representations compatible with spin ($\Gamma_{7,8,9}$) in this group are doubly-degenerate\cite{Koster.book.1963}, there is no spin-splitting along $k_z$.

In WZ, the spin splitting effects near zone center can be described with an effective one band Hamiltonian \cite{Casella1960,Hopfield1961,LewYanVoon1996}
\begin{equation}
H({\bf k}) \propto \frac{k_x^2+k_y^2}{2m_{\perp}}+\frac{k_z^2}{2m_{||}} + \displaystyle\sum_{n=0}^\infty \alpha_n[\sigma_xk_y+\sigma_yk_x]^{(2n+1)}
\label{WZ:H_k}
\end{equation}
where  $\alpha_n$ are constants. This effective Hamiltonian is invariant with respect to $C_{6v}$ for all odd powers of $k$. The coefficients of the linear and cubic Dresselhaus spin-splitting terms can be obtained by expanding Eq. \ref{WZ:H_k} up to $n=1$. Near the crossing, the difference in energies between the spin-split bands in Fig.~\ref{fig:WZ_spin}(a) has the form \cite{LewYanVoon1996}
\begin{equation}
|E_{i,\uparrow}(k)-E_{i,\downarrow}(k)|=2\zeta_1^ik + \zeta_3^i k^3
\label{WZ:spinfit}
\end{equation}
where  $\zeta_1^i$ and $\zeta_3^i$ are the linear and cubic Dresselhaus spin-splitting coefficients respectively for band $i$.  These coefficient characterize the spin spin splitting, and may be determined from the computed band structure by fitting a function of the above form. While the existence of a spin splitting in a particular direction is determined by symmetry considerations\cite{Casella1959}, the magnitude can vary considerably. Although both ZB and WZ crystals lack inversion symmetry, spin splitting effects are much more prominent in WZ due to its lower crystal symmetry \cite{Birman1959b,Hopfield1961b,Adler1962,Hummer1978,Broser1979,Koteles1980,Shigenari1982,LewYanVoon1996,Culcer2005PRB,Wang2007APL,Goano2007JAP,Fu2008JAP}, which can be of particular use in spintronic devices.

\section{Method} \label{sec:PP-method}
\subsection{Pseudopotentials}
Our band structures are computed using the empirical pseudopotential method of Cohen
and Chelikowsky \cite{Cohen.book} with a model potential applicable
to both ZB and WZ structures.
Pseudopotentials exploit the fact that the electronic wave function may be separated into the sum of a rapidly oscillating part near the atomic cores and a slowly varying piece.
The pseudopotential approach relies on the assumption that the core electrons are frozen and that the valence electrons move in a weak single-electron potential making the true atomic wave function orthogonal to the core states.
The pseudo wave equation is then
\begin{eqnarray}
    \left(\frac{p^2}{2m}+V_{pp}\right)|\phi\rangle = E|\phi\rangle
    \label{PPHo}
\end{eqnarray}
where $|\phi\rangle$ is the smoothly varying pseudo wave function, and  $V_{pp}$ is the pseudopotential which includes a repulsive core to
partially cancel the deep potential  near the atomic core\cite{Phillips1959}.
By taking $V_{pp}$ to be a local pseudopotential  which is a function only of position, it may be expanded in terms of reciprocal lattice vectors,  $\mathbf G$,  as
\begin{eqnarray}
 V_{pp}({\mathbf r}) &=& \displaystyle \sum_{{\bf G},\alpha} V_\alpha^{FF} ({\bf G})S_\alpha({\bf
G})e^{i\bf G\cdot r} \\
S_\alpha({\mathbf G})&=&\frac{1}{N_\alpha}\sum_{i=1}^{N_\alpha}
e^{-i\mathbf{G}\cdot\boldsymbol \tau_{\alpha,i}}
 \label{EMP2}
\end{eqnarray}
where $\alpha$ labels the atom type, $V^{FF}_\alpha({\bf G}) $ is the form factor, $S_\alpha({\mathbf G})$ is the structure factor, $N_\alpha$ is
the number of atoms per unit cell of type $\alpha$, and $\boldsymbol  \tau_{\alpha,i}$ is the position of atom number $i$ of type $\alpha$.

For binary compounds it is convenient to separate the pseudopotential into symmetric (S) and antisymmetric (A) parts as
\begin{eqnarray}
\langle\mathbf{G'}|{V}_{pp}|\mathbf{G}\rangle &=& V_S(\mathbf{G'}-\mathbf{G})S_S(\mathbf{G'}-\mathbf{G})\nonumber\\
&+& iV_A(\mathbf{G'}-\mathbf{G})S_A(\mathbf{G'}-\mathbf{G}).
\label{Ho.EPM}
\end{eqnarray}
The symmetric and antisymmetric structure factors are given by
\begin{eqnarray}
S_S(\mathbf{G})&=&\frac{1}{N}\displaystyle\sum_{j}\exp(-i\mathbf{G}\cdot{\boldsymbol\tau}_j) \label{SF1}\\
S_A(\mathbf{G})&=&\frac{-i}{N}\displaystyle\sum_{j}{P_j}\exp(-i\mathbf{G}\cdot{\boldsymbol\tau}_j) \label{SF2}
\end{eqnarray}
where $N$ is the number of atoms per unit cell and $P_j=+1$ for one type of atom and $-1$ for the other type. The
symmetric and antisymmetric form factors, $V_A({\bf G})$ and $V_S({\bf G})$,  are obtained from the sum and difference of the spherically symmetric anion and cation potentials.

There are several approaches to calculating $V_{FF}$ \cite{Cohen.book,Heine.book}.
In the empirical pseudopotential approach used here, $V_S({\bf G})$ and  $V_A({\bf G})$ are adjusted to fit the calculated energy spectrum to experimentally determined energies at band extrema.
Model potentials that yield an accurate band structure of a known polytype should reliably predict the band  structure for the unknown polytype if the two crystal structures are similar.
To compute the WZ band structure using pseudopotentials obtained from ZB requires $V_S({\bf G})$ and  $V_A({\bf G})$ to be continuous functions that can be evaluated at any value of $\bf G$.
A wide variety model potentials have been used\cite{Heine.book,Xia1988,Yeh1994b,Pugh1999,Fan2006} and we use potentials of the form
\begin{eqnarray}
V_S({\bf G})&=&\frac{x_1G - x_2}{\exp({x_3G^2 + x_4})+1} \label{FF.func1} \\
V_A({\bf G})&=&\left( x'_1G^2 + x'_2 \right) \exp({x'_3G^2 + x'_4})  \label{FF.func2}
\end{eqnarray}
where $G=|\bf{G}|$, and  the parameters $x_{j}$ and $x'_j$ are obtained for each material by fitting to the ZB band structure.
\subsection{Spin-Orbit Interactions}
We include the spin-orbit coupling, given by
\begin{eqnarray}
 H_{so}=
\frac{\hbar^2}{4m^2c^2}\left(\bf\nabla{\bf\it{V(\bf{r})}}
\times\bf{p}\right)\cdot{\boldsymbol \sigma} \label{Hso}
\end{eqnarray}
using the method of Weisz\cite{Weisz1966}.
Eq.~\ref{Hso} cannot be used directly  since spin-orbit coupling involve the core states which are omitted from the pseudo wave equation as a result of orthogonalization.
Instead, by returning to the original Schr\"odinger equation we may expand in terms of the core states  $|\phi_c\rangle$ to obtain \cite{Weisz1966, Saravia1968}
\begin{eqnarray}
\langle\phi'| V_{so}|\phi\rangle \approx
\sum_{c,c'}\langle\phi'|\phi_{c'}\rangle\langle\phi_{c'}|{H}_{so}|\phi_c\rangle\langle\phi_c|\phi\rangle .
\label{Vso2}
\end{eqnarray}
In this way the matrix elements of $H_{so}$ for core states may be parameterized and then fit to experiment in the same way as the pseudopotential form factors.
By expanding $|\phi\rangle$ in Bloch functions, and expressing the
core states in terms of its constituent atomic radial wave functions
and respective spherical harmonics, Eq.~\ref{Vso2} can be recast
as \cite{Weisz1966}
\begin{eqnarray}
   \langle \mathbf{K'},s'| {V}_{so}|\mathbf{K},s\rangle&=&(\mathbf{K'}\times\mathbf{K})\cdot\langle{s'}|{\boldsymbol\sigma}|{s}\rangle \nonumber \\
   && \displaystyle\sum_{l} \lambda_l ~{\rm\it{P}}_l'(\cos\theta_{\mathbf { K'\cdot K} })S(\bf
   K'-K) \nonumber \\
\label{SOC1}
\end{eqnarray}
where $P'_l$ is the derivative of the Legendre polynomial,
$\boldsymbol \sigma$s are the Pauli matrices,
 $\bf{K=G+k}$,
and  $\theta$ is the angle between ${\bf K}$ and ${\bf K}'$.
The coefficient   $\lambda_l$ is given in terms of the core wave functions by
\begin{equation}
\lambda_l = \mu\beta_{nl}({\mathbf K'})\beta_{nl}({\mathbf K})
\end{equation}
\begin{equation}
\beta_{nl}(K) = C
{\displaystyle{\int_0^\infty}}i^l\sqrt{4\pi(2l+1)}j_{nl}(Kr)R_{nl}(r)r^2dr
\label{beta}
\end{equation}
where $C$ is a normalization constant such that $\beta_{nl}(K)/K$
approaches unity in the limit $K$ goes to zero and $n$ is the
principal quantum number for the core state being considered. For
III-V semiconductors, it is not required to expand Eq.~\ref{SOC1}
beyond $l=2$ since they do not have core shells filled beyond
$d$-orbitals. Expanding Eq.~\ref{SOC1} up to $l=2$, the matrix
elements for the spin-orbit coupling in a binary compound are
\begin{eqnarray}
& \langle \mathbf{K'},s'| {V}_{so}|\mathbf{K},s\rangle=
-i(\mathbf{\hat{K}'}\times\mathbf{\hat{K}})\cdot\langle{s'}| {\boldsymbol \sigma}|{s}\rangle \nonumber \\
&\left[       \left( \lambda^S_p+ \lambda^S_d \, \mathbf{\hat{K}'}\cdot\mathbf{\hat{K}} \right)  S_S(\mathbf{G'}-\mathbf{G}) \right. \nonumber \\
&{}+ \left. \left(\lambda^A_p+  \lambda^A_d \, \mathbf{\hat{K'}}\cdot\mathbf{\hat{K}} \right)   S_A(\mathbf{G'}-\mathbf{G})  \right]
\label{Vso.EMP1}
\end{eqnarray}
\begin{eqnarray}
\lambda^S_l &=&  \left( \lambda_l^{(1)}+\lambda_l^{(2)} \right)/2\\
\lambda^A_l &=&  \left( \lambda_l^{(1)}-\lambda_l^{(2)} \right)/2 \\
\lambda_l^{(1)} &=& \mu_l~\beta^{(1)}_{nl}({\bf{K}}_i)~\beta^{(1)}_{nl}({\bf{K}}_j)\\
\lambda_l^{(2)} &=& \gamma_l~\mu_l~\beta^{(2)}_{nl}({\bf{K}}_i)~\beta^{(2)}_{nl}({\bf{K}}_j)
  \label{Vso.EMP2}
\end{eqnarray}
where the superscript (1),(2) specifies which atom, the coefficient $\mu_l$ is an empirically adjusted parameter and $\gamma_l$ is the
ratio of the anion to cation spin-orbit splitting energies for a given core state\cite{Chelikowsky1976}.
The overlap integral, $\beta_{nl}$, is constructed from the atomic core wave functions using Eq.~\ref{beta}.
The radial part of the core wave function, $R_{nl}$, is an approximate Hartree-Fock solution taken from Herman-Skillman
tables\cite{Herm.Skill}.
For Ga, In, As and Sb, terms up to $l=2$ in
Eq.~\ref{Vso.EMP1} are included while Al and P only go up to $l=1$ since they do
not have
valence d-shells.
With the inclusion of Eq.~\ref{Vso.EMP1} the total pseudopotential Hamiltonian becomes
\begin{eqnarray}
{H} = \frac{-\hbar^2\mathbf{K}^2}{2m} + {V}_{pp} + {V}_{so}.
\label{HoVso}
\end{eqnarray}
\subsection{Fitting}
The pseudopotential parameters $x_i$, $x'_i$, and $\mu_i$ in Eqs.~\ref{FF.func1} and \ref{FF.func2} were determined by fitting the band structure obtained from the Hamiltonian of Eq.~\ref{HoVso} to experimental energies of the band extrema of ZB materials.
The Hamiltonian was evaluated in a plane wave basis with a cutoff of $|{\bf G}| \le 32 \pi / a$, and
for each value of $\bf k$ the Hamiltonian was diagonalized to give energies to be fit to the experimental target values.
The fitting was accomplished by minimizing the error function
\begin{equation}
F = \displaystyle\sum_i W_i \frac{ \left[ E_i(calculated)-E_i(target) \right]^2}{E^2_i(target)}
\label{err}
\end{equation}
where the sum over $i$ ranges over the targeted
energies $E_i$, and $W_i$ are weighting factors adjusted to speed convergence.
Seven energies were used as fitting parameters, (all with respect to $E^{\Gamma}_{8v}=0$):
$E^{\Gamma}_{6c}, E^{\Gamma}_{7v}, E^{X}_{6c},
E^{L}_{6c}, E_{v6}^{\Gamma}, E_{c7}^{\Gamma}, E_{c8}^{\Gamma}$.
The targeted values of $E^\Gamma_{6c}$, $E^{X}_{6c}$, $E^L_{6c}$
and $E^\Gamma_{7v}$ were taken from Ref.
\onlinecite{Vurgaftman2001} while the higher transition energies, were taken from Ref. \onlinecite{Madelung}.
In addition constraints were imposed to ensure
the correct band ordering of valence states by forcing the third and
fourth (spin-degenerate) valence band states to have $\Gamma_7$
symmetry.
$F$ was minimized with respect to  $x_i$, $x'_i$, and $\mu_i$ using Powell's method, with the local pseudopotential form factors from Ref. \onlinecite{Cohen1966} used as an initial starting point.
Slightly different initial values were used as a check that the solution did not converge to a spurious local minimum since Powell's method finds the local minimum.
In those cases where the value of $F$ was large, indicating a poor fit, the weights $W_j$ were  adjusted to climb out of the local minimum in which the algorithm was trapped, and the minimization algorithm was continued.
\subsection{Transferable Pseudopotentials}
Once the form factors have been determined for the ZB polytype, they may be transfered to the WZ structure by centering the spherically symmetric atomic pseudopotentials on the positions of the ions in the WZ form.
The transferability depends on the similarities of the crystal structures, and for sufficiently dissimilar polytypes one
would expect the method to fail.
Fortunately, as discussed in section \ref{sec:crystalStructure} ZB and WZ are very similar, as the WZ crystal structure can be thought of as a variation of ZB with  the same local structure, but a slightly different long range structure.
It is important to note that all the parameters were fit independently for each material, and the pseudopotentials were transferred between polytypes with the same binary composition.
Since the WZ primitive cell has four atoms (rather than the two of ZB), its structure factor
contain more terms.
Substituting the atomic positions  in section \ref{sec:crystalStructure} into Eqs.~\ref{SF1} and \ref{SF2} the WZ structure factors are
\begin{eqnarray}
S_S=\frac{1}{4}\left( 1+e^{-iG_3uc}\right) \left( 1+e^{-iG_2a/\sqrt{3}+G_3c/2}\right) \\
S_A= \frac{1}{4}\left( 1-e^{-iG_3uc}\right) \left(1+e^{-iG_2a/\sqrt{3}+G_3c/2}\right)
\label{SF.WZ}
\end{eqnarray}
 where $G_j$ ($j=1,2,3$) are the components of the reciprocal lattice vector $\bf{G}$.

\section{Results}\label{sec:results}
        \subsection{Calculated III-V Zinc Blende band structures}
The pseudopotential parameters determined by fitting to the zincblende band structures are given in Table \ref{tab:ZB-FF}, including the spin-orbit parameters.
The accuracy of the results may be gauged by Table \ref{tab:ZB-conv}, which gives the ratio of each band energy to the experimental value to which it was fit\cite{Vurgaftman2001}.
We see that the results agree with experiment to within $1\%$ for all but $E_{6v}^\Gamma$.
Even for $E_{6v}^\Gamma$, the deviation from the experimental value is greater than $10\%$ only for InSb.
We have also made a similar comparison of the effective masses and Luttinger parameters\cite{Vurgaftman2001}.
These values differ from the energy ratios in that the masses were not fit to experimental data.
The masses were determined by doing a quadratic fit to the band extremum, with the Luttinger parameters determined from
\begin{eqnarray}
    \left(\frac{m^*}{m_{hh(lh)}}\right)^{[001]} = \gamma_1 \mp 2\gamma_2 \nonumber \\
    \left(\frac{m^*}{m_{hh(lh)}}\right)^{[111]} = \gamma_1 \pm 2\gamma_3
\label{Luttinger}
\end{eqnarray}
It should be noted that  several of the masses listed in
Ref. \onlinecite{Vurgaftman2001} are either obtained theoretically
or have large experimental uncertainties. For example, only
theoretically calculated effective masses are available for AlP
and for the valence band of AlSb. To the best of our knowledge,
experimental results are not available for these compounds.
In the case of GaP the conduction band effective mass is extrapolated from its ternary alloy.
For InAs, there is great experimental uncertainty about the heavy and light hole masses
\cite{Vurgaftman2001}.
Even though we have omitted nonlocal corrections not associated with spin-orbit coupling, our effective masses are in very good agreement for compounds containing lighter elements.
\begin{center}
\begin{table*}[ht]
\begin{tabular}{c|c c c c c c c c c c}
 \hline
 \hline
{Material} & $x_1$ & $x_2$  &  $x_3$  &  $x_4$  &  $x_1'$  & $x_2'$ & $x_3'$ & $x_4'$ & $\mu_1$ & $\mu_2$ \\
\hline
AlP     &   0.083  &  -0.579  &  0.031  &  -2.586  &  -0.11   &  0.9    &  -0.061  &  -1.178  &  0.012 & 0  \\
AlAs    &   0.062  &  -0.459  &  0.027  &  -2.629  &  -0.041  &  1.003  &  -0.056  &  -1.693  &  $5.28 \times 10^{-3}$  & $5 \times 10^{-5}$ \\
AlSb    &   0.06   &  -0.412  &  0.032  &  -2.548  &  -0.09   &  0.17   &  -0.051  &  -2.098  &  $7.26 \times 10^{-3}$  & $2.96 \times 10^{-4}$ \\
GaP     &   0.085  &  -0.457  &  0.04   &  -2.566  &  -0.351  &  5.165  &  -0.205  &  0.339   &  0.385                  & $-2.3 \times 10^{-3}$ \\
GaAs    &   0.058  &  -0.467  &  0.023  &  -2.583  &  -0.063  &  1.091  &  -0.074  &  -1.298  &  0.052                  & $8.3 \times 10^{-6}$ \\
GaSb    &   0.042  &  -0.343  &  0.022  &  -2.584  &  -0.009  &  0.618  &  -0.043  &  -2.233  &  0.056                  & $2.78 \times 10^{-5}$ \\
InP     &   0.049  &  -0.385  &  0.027  &  -2.602  &  0       &  0.847  &  -0.059  &  -1.654  &  0.243                  & $-1.09 \times 10^{-3}$ \\
InAs    &   0.036  &  -0.298  &  0.033  &  -2.615  &  -0.011  &  1.359  &  -0.121  &  -1.124  &  0.082                  & $2.603 \times 10^{-5}$\\
InSb    &   0.022  &  -0.174  & 0.023   &  -2.42   &  -0.012  &  1.158  &  -0.082  &  -1.363  &  0.085                  & $5.7 \times 10^{-5}$ \\
\hline
\end{tabular}
\caption { Fitting parameters for symmetric and antisymmetric form factors, where the form factors are in units of Ry.
$\mu_1$ and $\mu_2$ are the fitting parameters for the spin-orbit coupling.
All fitting coefficients are rounded off to the third decimal place. }
\label{tab:ZB-FF}
\end{table*}
\end{center}
\begin{center}
\begin{table*}[ht]
\begin{tabular}{c|c c c c c c c c |c c c c c}
 \hline
 \hline
{Material} & $E^{\Gamma}_{6c}$ & $\Delta_{so}$  &  $E^{L}_{6c}$  &  $E^{X}_{6c}$  &  $E^{\Gamma}_{7c}$  & $E^{\Gamma}_{8c}$ & $\Delta_{so}'$ & $E^{\Gamma}_{6v}$  & $m_c$ & $m_{so}$  &  $\gamma_1$  &  $\gamma_2$  &  $\gamma_3$\\
\hline
AlP     &   1.00   &   1.00    &   1.00   &   1.00     & -      &    -       &   -     &   1.03 &   0.72   &   1.06    &   0.92   &   0.99     &   0.95  \\
AlAs    &   1.00   &   1.00    &   1.01   &   1.00     &   1.01   &     1.01 &   1.00  &   0.99 &   0.99   &   1.07    &   0.95   &   0.84     &   0.96 \\
AlSb    &   1.00   &   1.00    &   1.00   &   1.00     &   1.00   &     1.00 &   1.00  &   0.97 &   1.04   &   1.30    &   0.67   &   0.66     &   0.70 \\
GaP     &   1.00   &   1.00    &   1.00   &   1.00     &   1.00   &     1.00 &   1.00  &   0.97 &   1.00   &   1.02    &   0.96   &   1.66     &   1.20   \\
GaAs    &   1.00   &   1.00    &   1.00   &   1.00     &   1.00   &     1.00 &   1.00  &   0.95 &   1.16   &   1.19    &   0.81   &   0.90     &   0.83 \\
GaSb    &   1.00   &   1.00    &   1.01   &   1.00     &   1.00   &     1.00 &   1.00  &   0.94 &   1.47   &   1.56    &   0.63   &   0.61     &   0.62 \\
InP     &   1.00   &   1.00    &   1.00   &   1.00     &   1.00   &     1.00 &   1.00  &   1.02 &   1.08   &   1.03    &   0.96   &   0.98     &   1.00 \\
InAs    &   1.00   &   1.00    &   1.00   &   1.01     &   1.00   &     1.00 &   1.00  &   0.95 &   1.36   &   1.05    &   0.61   &   0.60     &   0.62 \\
InSb    &   1.00   &   1.00    &   1.01   &   1.00     &   1.00   &     1.00 &   1.00  &   0.84 &   1.85   &   1.73    &   0.55   &   0.54     &   0.56 \\
\hline
\end{tabular}
\caption {Ratios of calculated/targeted band transition energies at
various high symmetry points for nine III-V zincblende
semiconductors. All transition energies are referenced to the top of
valence band, ($E^{\Gamma}_{8v}$). The spin-orbit
energies are $\Delta_{so}=E^{\Gamma}_{8v} -
E^{\Gamma}_{7v}$ and $\Delta_{so}'=E^{\Gamma}_{8c} -
E^{\Gamma}_{7c}$.
In addition, the effective for the conduction band ($m_c$),
split-off bands($m_{so}$), and heavy and light holes
are compared to those from Ref. \onlinecite{Vurgaftman2001}.
The heavy
and light hole masses are expressed in terms of Luttinger parameters
$\gamma_{1},\gamma_2$ and $\gamma_3$ (see Eq.~\ref{Luttinger}).
Note, that the effective masses were not set as targets for fitting
the form factors.}
\label{tab:ZB-conv}
\end{table*}
\end{center}

       \subsection{Predicted III-V Wurtzite band structures}
\begin{figure}
  \includegraphics[width=1\columnwidth]{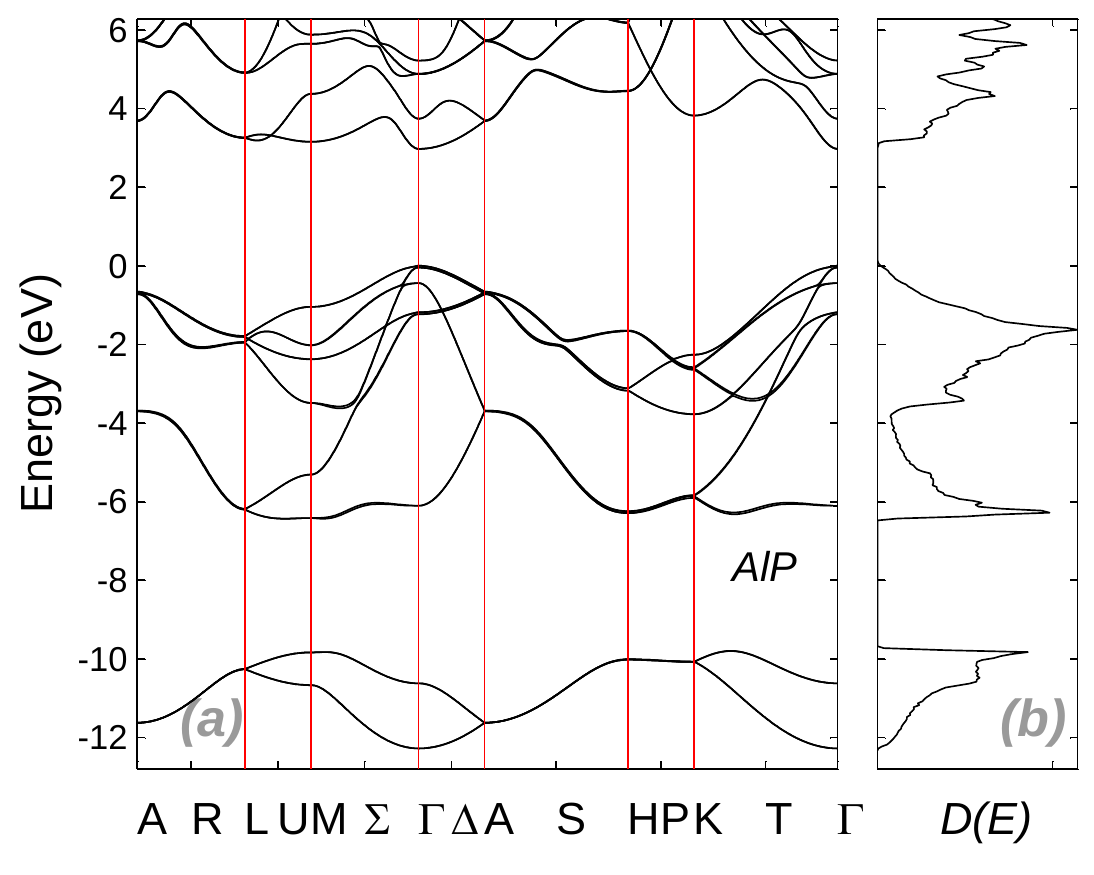}
  \caption{ ({\bf a}) Calculated low temperature band structure for AlP in WZ phase. ({\bf b}) Calculated DOS.}
  \label{fig:WZ-AlP}
\end{figure}

\begin{figure}
  \includegraphics[width=1\columnwidth]{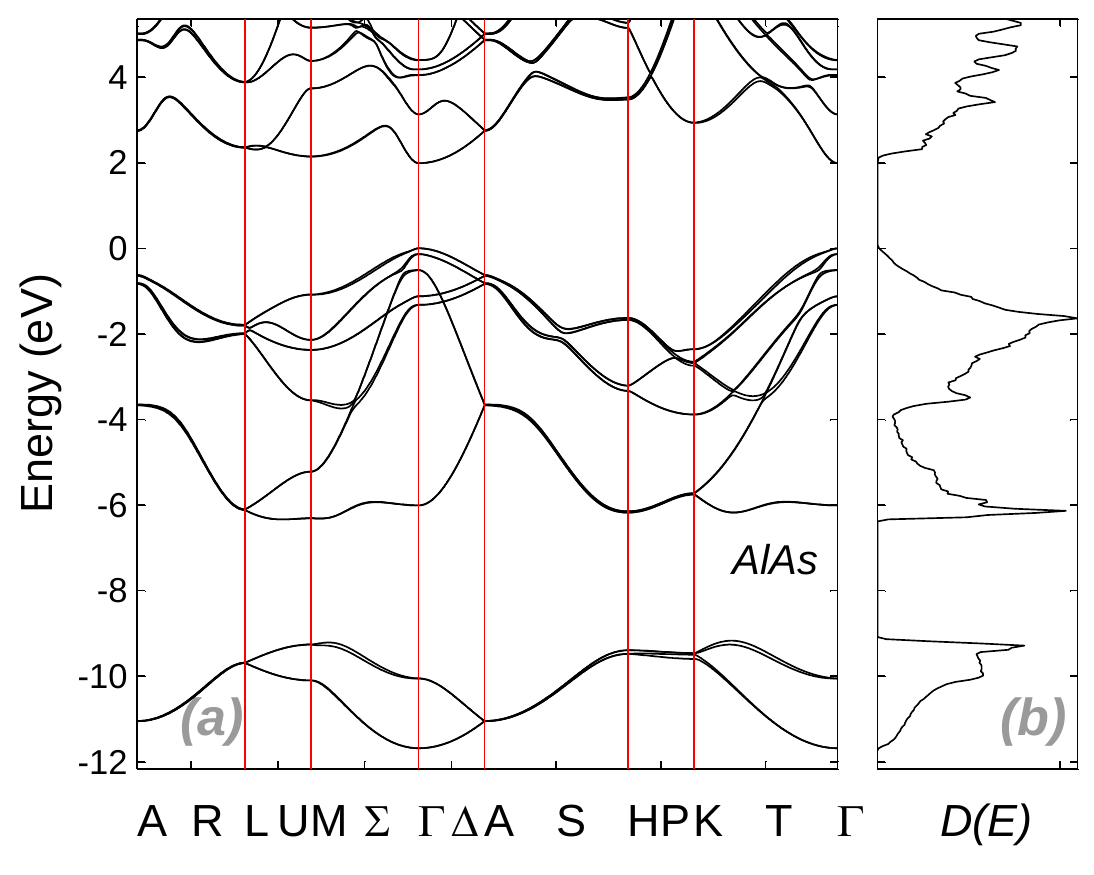}
  \caption{ ({\bf a}) Calculated low temperature band structure for AlAs in WZ phase. ({\bf b}) Calculated DOS.}
  \label{fig:WZ-AlAs}
\end{figure}

\begin{figure}
  \includegraphics[width=1\columnwidth]{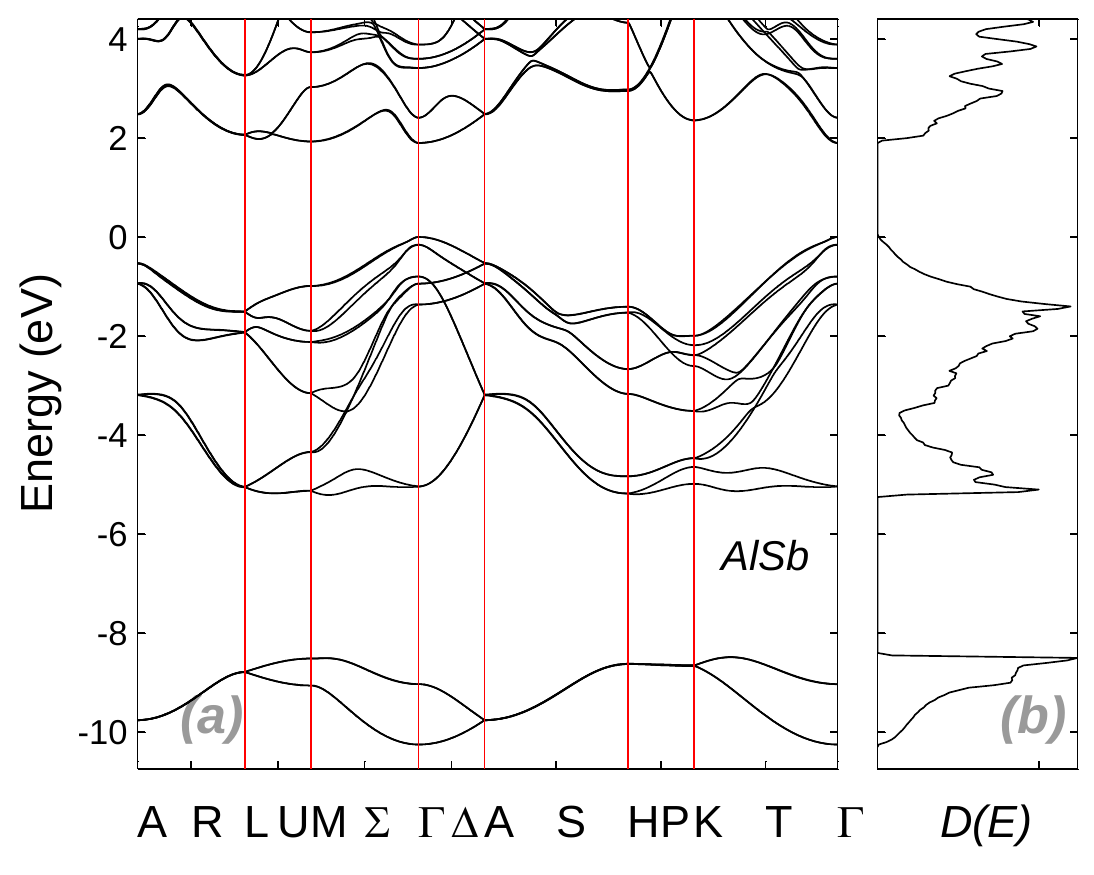}
  \caption{ ({\bf a}) Calculated low temperature band structure for AlSb in WZ phase. ({\bf b}) Calculated DOS.}
  \label{fig:WZ-AlSb}
\end{figure}

\begin{figure}
  \includegraphics[width=1\columnwidth]{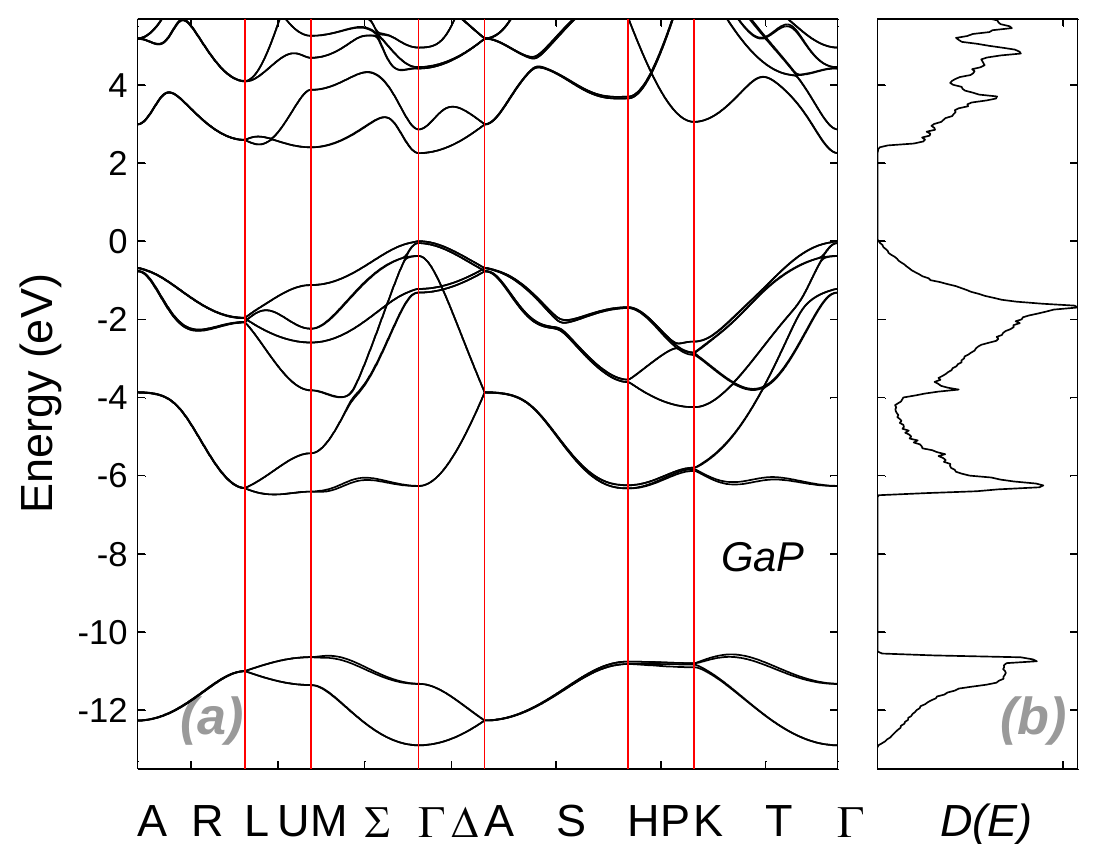}
  \caption{ ({\bf a}) Calculated low temperature band structure for GaP in WZ phase. ({\bf b}) Calculated DOS.}
  \label{fig:WZ-GaP}
\end{figure}

\begin{figure}
  \includegraphics[width=1\columnwidth]{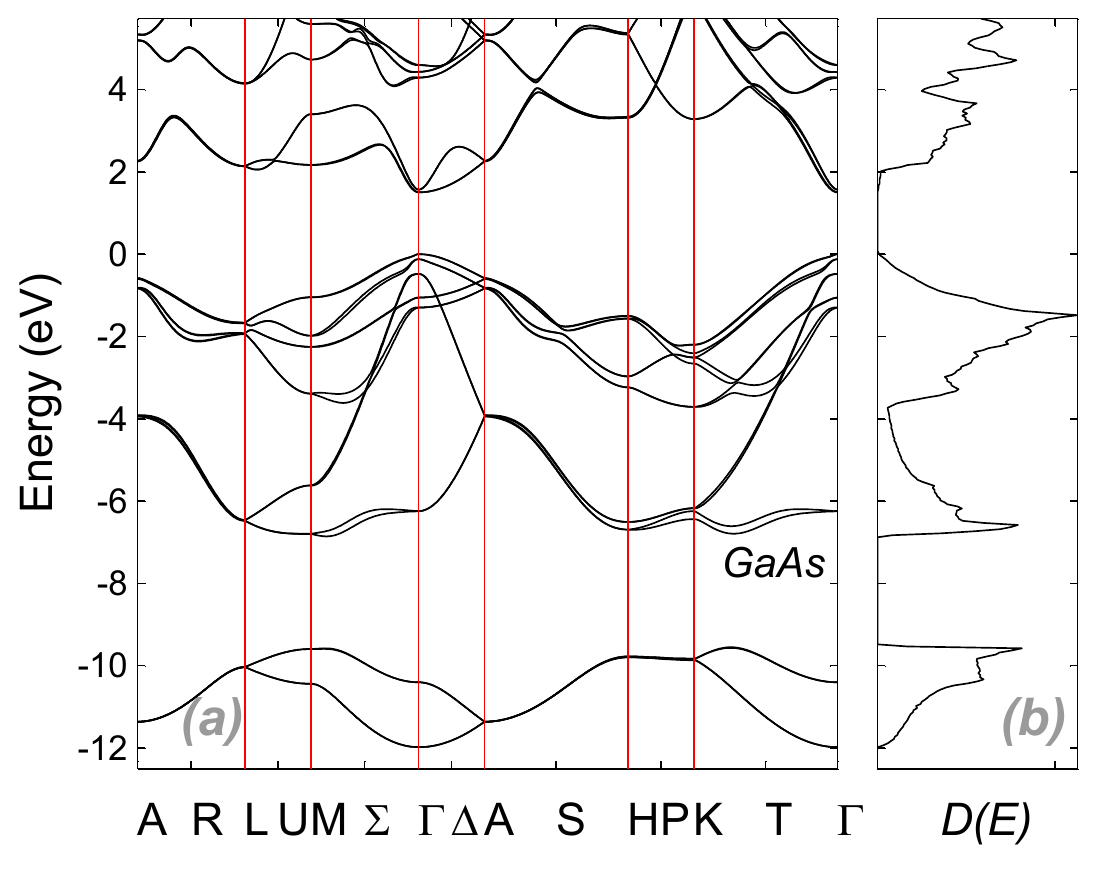}
  \caption{ ({\bf a}) Calculated low temperature band structure for GaAs in WZ phase. ({\bf b}) Calculated DOS.}
  \label{fig:WZ-GaAs}
\end{figure}

\begin{figure}
  \includegraphics[width=1\columnwidth]{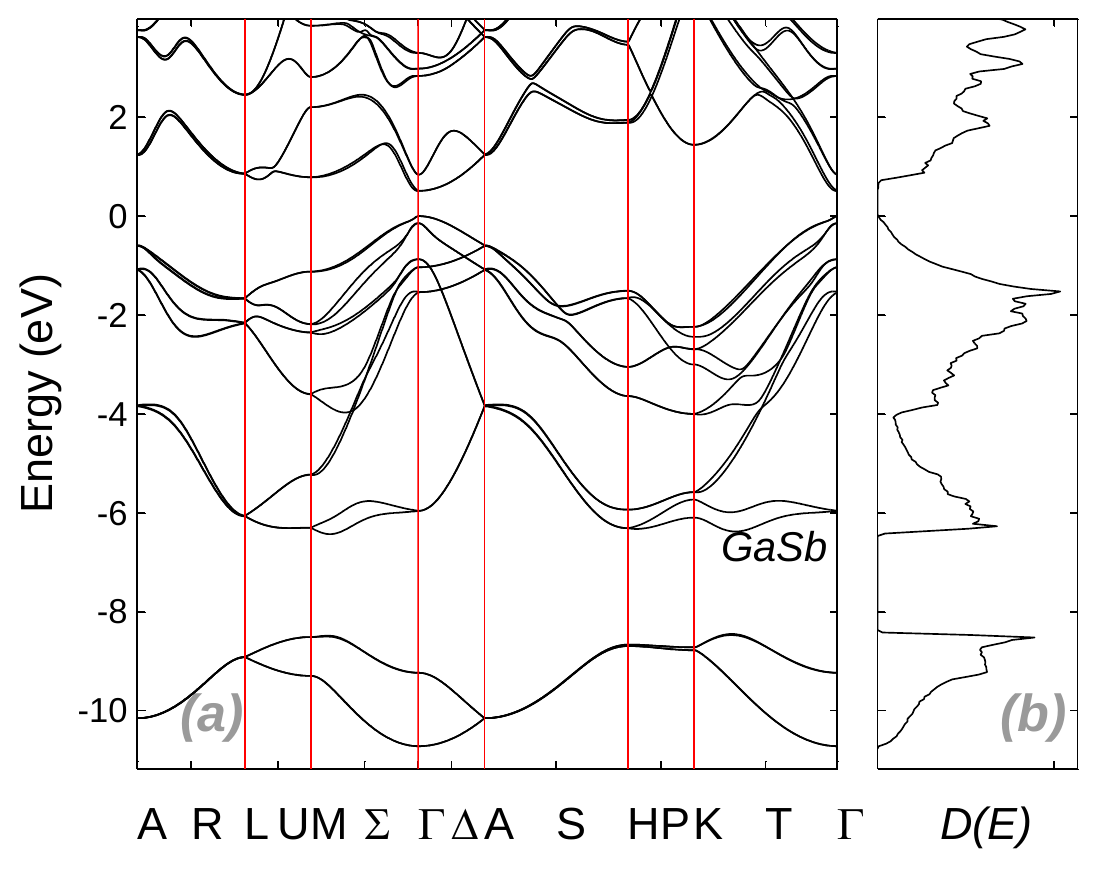}
  \caption{ ({\bf a}) Calculated low temperature band structure for GaSb in WZ phase. ({\bf b}) Calculated DOS.}
  \label{fig:WZ-GaSb}
\end{figure}

\begin{figure}
  \includegraphics[width=1\columnwidth]{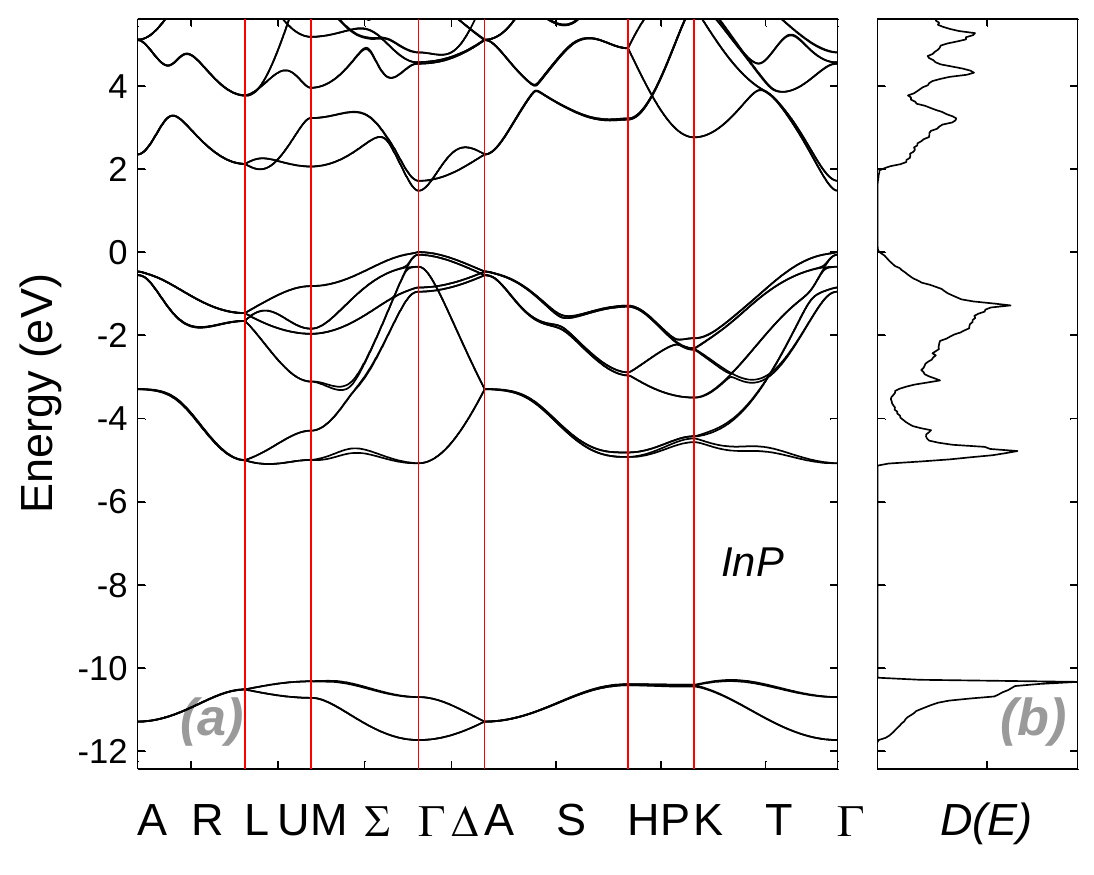}
  \caption{({\bf a}) Calculated low temperature band structure for InP in wurtzite phase. ({\bf b}) Calculated DOS.}
  \label{fig:WZ-InP}
\end{figure}

\begin{figure}
  \includegraphics[width=1\columnwidth]{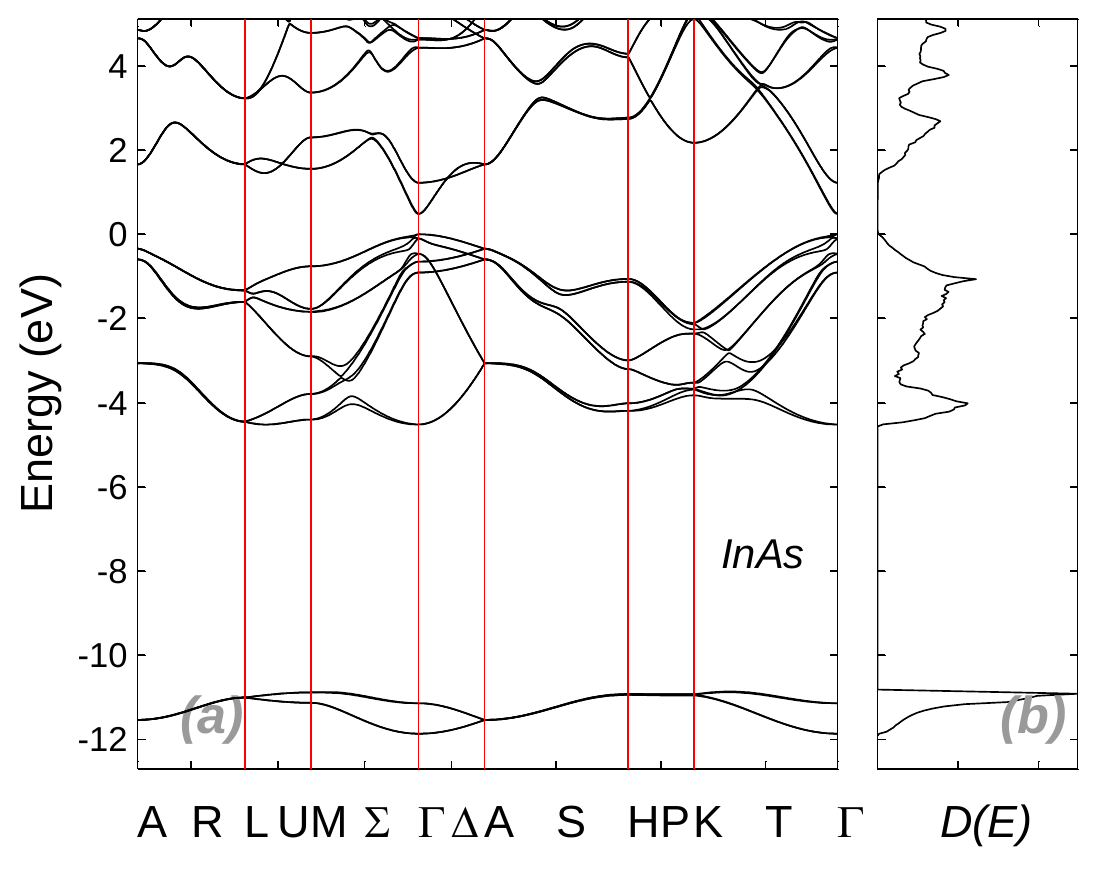}
  \caption{({\bf a}) Calculated low temperature band structure for InAs in wurtzite phase. ({\bf b}) Calculated DOS.}
  \label{fig:WZ-InAs}
\end{figure}

\begin{figure}
  \includegraphics[width=1\columnwidth]{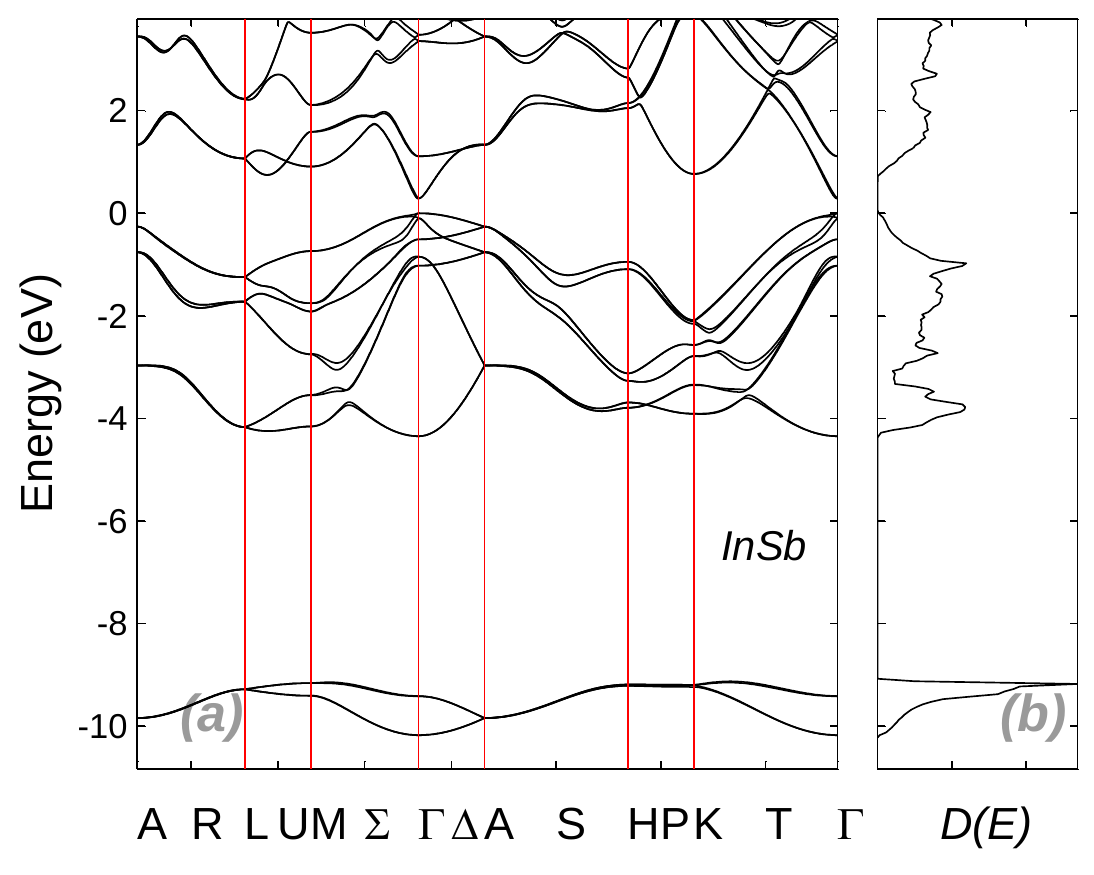}
  \caption{({\bf a}) Calculated low temperature band structure for InSb in wurtzite phase. ({\bf b}) Calculated DOS.}
  \label{fig:WZ-InSb}
\end{figure}

The calculated band structure and the corresponding density of states(DOS) for each of the nine III-V semiconductors in WZ phase are shown in Fig. \ref{fig:WZ-AlP}-\ref{fig:WZ-InSb}. The electronic band structures are calculated in the irreducible wedge of the Brillouin zone (Fig. \ref{fig:ZBvsWZ}(d)). It should be noted that the band structure of WZ is more complicated than that of ZB due to it lower crystal symmetry and has roughly twice as many bands over a given energy range. The irreducible representations of the zone center states was determined by transforming the pseudo wave functions under the symmetry operations of the respective crystallographic point group.

Tables \ref{tab:WZ-vals1}-\ref{tab:WZ-vals3} list the calculated zone center energies of the band extrema, effective masses for ${\bf k}$ parallel and perpendicular to the $c$ axis, and the linear and cubic Dresselhaus spin splitting coefficients $\zeta_1$ and $\zeta_2$. We calculated the Dresselhaus coefficients by fitting a function of the above form in Eq.~\ref{WZ:spinfit} to the calculated band structures. These parameters may be used in constructing $\bf k{\cdot}p$ WZ Hamiltonians for nanostructure calculations
\cite{Foley1986,Chuang1996PRB,Yeo1998,Sirenko1997,Xia1999,Mireles2000,Rodina2001,Beresford2004}.
We have also complied a summary (Table \ref{tab:WZ-compact}) listing the energy differences most important for nanostructures.

Foremost is the band gap and the irreducible representation of the conduction band minimum. In our calculations it is seen that all of the materials containing Al or Ga have $\Gamma_{8}$ conduction band minima in the WZ phase, whereas all of the materials containing In have $\Gamma_{7}$ conduction band minima. We also give the  spin-orbit  energy, $\Delta_{so}$, the crystal field splitting, $\Delta_{cr}$, and the offset between the valence band edge
of each polytypes.
$\Delta_{so}$ and $\Delta_{cr}$ are extracted using the quasi cubic approximation which assumes WZ to be equivalent to
$[111]$-strained ZB \cite{BirPikus,Chuang1996PRB}, with $\Delta_{so}$ and $\Delta_{cr}$  related to the $\Gamma_{7v}$ hole energies by
\begin{eqnarray}
E(\Gamma_{7v}^{1,2}) - E(\Gamma_{9v}) = -\frac{\Delta_{so}+\Delta_{cr}}{2}\nonumber\\
\pm\frac{\sqrt{(\Delta_{so}+\Delta_{cr})^2-u^{-1}\Delta_{so}\Delta_{cr}}}{2}
\label{WZ:CrSo}
\end{eqnarray}
where $u$ is a parameter relating the lattice constants, $\sqrt{u}=a/c$.  For the ideal WZ structure assumed here,  and is $3/8$ for an ideal WZ structure.

In ZB phase, AlP, AlAs, AlSb and GaP are indirect gap semiconductors with their conduction band minima ordered $X$, $L$, $\Gamma$. Our calculations show that in all of the indirect gap ZB semiconductors become direct gap WZ materials with $\Gamma_{8}$ conduction band minima.
Previous LDA calculations obtained indirect gaps for AlP and AlAs in WZ phase, with the conduction band minimum at $M$ \cite{Yeh1994, Murayama1994}.
Although this is not the case in our results, we do find the $M$ valley conduction band minimum only slightly above the $\Gamma$ minima (182 meV for AlP and 157 meV for AlAs). The same LDA calculations also predicted AlSb  and GaP to have $\Gamma_{8}$ conduction band minima, in agreement with our results\cite{Murayama1994}.
The direct gaps of AlP and AlSb  in WZ phase are larger than the indirect  ($X$-valley) gaps of their respective ZB counterparts.
Whereas the direct ($\Gamma_8$) gaps of AlAs and GaP in WZ phase are smaller than the indirect gaps of their ZB counterparts.

Although GaAs and GaSb have direct gaps in the ZB phase, for both materials the zone-folded $L$ valley in WZ is lower, giving a  $\Gamma_{8}$ minimum.
For WZ, GaAs  has a band gap ($\approx 1.5~\rm  eV$) which is only slightly smaller than that of its ZB polytype, while GaSb has a significantly smaller gap ($\approx 0.53~\rm eV$) in WZ phase. In both cases this is due to $E_{L1c}$ being close to $E_{\Gamma1c}$. This behavior is most apparent for GaSb, hence our prediction for GaSb is consistent with that of Ref. \onlinecite{Yeh1994} and Ref. \onlinecite{Murayama1994}. In light of recent experimental results, the case of WZ GaAs will be discussed in greater detail below. All three Indium containing compounds InP, InAs and InSb are direct gap semiconductors in WZ phase with $\Gamma_7$ conduction band minima and have higher band gaps than their respective direct gap ZB polytypes.

While little is known about the WZ polytypes, experimental data is available for the band gaps of GaAs, InAs and InP.
Table (\ref{tab:gap-compare}) compares our calculated band gaps for these three materials to experiment and {\it ab initio} calculations.
We have only tabulated their SX and GW results as they make different qualitative predictions as to whether the band gap of WZ GaAs is larger or smaller than for ZB.
Note that the values listed for the {\it ab initio} methods are not listed directly in the respective references.
As none of the directly obtained gaps from {\it ab initio} calculations are correct,  the values are obtained from the experimental ZB gap and the calculated percentage change between the two polytypes.
As can be seen, our results are in very good agreement with experiments.
More importantly, they are in agreement with the
experimental trends as to whether the WZ band gap is larger or smaller than the ZB gap.
This is most apparent in the case of GaAs, for which {\it ab initio} calculations (except the SX method\cite{Zanolli2007prb}) predict that GaAs should have a larger direct band gap in WZ phase than in ZB, in disagreement with experiment.
All of the low temperature experimental results show that for GaAs the WZ phase has a smaller band gap than ZB.
In addition, the experimental gaps are obtained from photoluminescence measurements on GaAs nanowires, which will be slightly larger due to confinement.

In the case of InP our results are in agreement with experiments as well as with the trends from LDA calculations.
Photocurrent spectroscopy measurements on $\rm InAs_{1-x}P_x$ nanowires have been extrapolated in $x$ to obtain a band gap for InP of $1.645~\rm eV$\cite{Tragardh2007}, which  is higher than valules from photoluminescence measurements.
The fact that the measured values are higher than our pseudopotential calculations may be du to confinement effects in the nanowire experiments.
Our calculated gap for InP is also in good agreement with the calculations of Zanolli {\it et al.} \cite{Zanolli2007prb}.

\begin{sidewaystable}
\begin{tabular}{||c|c|c|c|c|c||c|c|c|c|c|c||c|c|c|c|c|c||}
\hline \hline
\multicolumn{6}{||c|}{\bf AlP} & \multicolumn{6}{|c|}{\bf AlAs} & \multicolumn{6}{|c||}{\bf AlSb} \\
\hline
     $IR$ &  $E (eV)$ &  $m_{||}$ & $m_{\perp}$ & $\zeta_1 (eV \AA )$ & $\zeta_3 (eV \AA^3 )$ &      $IR$ &  $E (eV)$ &  $m_{||}$ & $m_{\perp}$ & $\zeta_1 (eV \AA )$ & $\zeta_3 (eV \AA^3 )$ &      $IR$ &  $E (eV)$ &  $m_{||}$ & $m_{\perp}$ & $\zeta_1 (eV\AA )$ & $\zeta_3 (eV \AA^3 )$ \\
\hline
$\Gamma_7$ &    -12.368 &      1.351 &      1.365 &      0.000 &    -22.061 & $\Gamma_7$ &    -11.763 &      1.307 &      1.318 &      0.000 &    -28.806 & $\Gamma_7$ &    -10.299 &      1.405 &      1.424 &      0.001 &   -164.203 \\

$\Gamma_8$ &    -10.710 &      0.550 &      1.598 &      0.001 &     -0.019 & $\Gamma_8$ &    -10.131 &      0.502 &      1.512 &      0.113 &     -0.143 & $\Gamma_8$ &     -9.078 &      0.629 &      1.820 &      0.021 &     -0.186 \\

$\Gamma_8$ &     -6.195 &      0.298 &      4.002 &      0.038 &      0.115 & $\Gamma_8$ &     -6.088 &      0.280 &      3.246 &      0.004 &      0.256 & $\Gamma_8$ &     -5.090 &      0.319 &      2.357 &      0.296 &      1.485 \\

$\Gamma_8$ &     -1.249 &      1.655 &      0.299 &      0.074 &     23.745 & $\Gamma_8$ &     -1.341 &      1.608 &      0.257 &      0.164 &      8.293 & $\Gamma_8$ &     -1.377 &      1.611 &      0.284 &      0.577 &     16.066 \\

$\Gamma_9$ &     -1.200 &      1.662 &      0.296 &      0.000 &     24.557 & $\Gamma_9$ &     -1.131 &      1.658 &      0.235 &      0.000 &      9.896 & $\Gamma_9$ &     -0.955 &      1.667 &      0.202 &      0.000 &     10.236 \\

$\Gamma_7$ &     -0.435 &      0.145 &      1.260 &      0.047 &    -11.829 & $\Gamma_7$ &     -0.518 &      0.150 &      0.837 &      0.140 &    -59.309 & $\Gamma_7$ &     -0.802 &      0.237 &      0.971 &      0.095 &   -234.420 \\

$\Gamma_7$ &     -0.044 &      0.931 &      0.253 &      0.052 &     25.634 & $\Gamma_7$ &     -0.139 &      0.478 &      0.259 &      0.236 &     -7.256 & $\Gamma_7$ &     -0.156 &      0.220 &      0.373 &      0.125 &     17.954 \\

$\Gamma_9$ &      0.000 &      0.972 &      0.248 &      0.000 &     25.032 & $\Gamma_9$ &      0.000 &      0.933 &      0.216 &      0.000 &     10.703 & $\Gamma_9$ &      0.000 &      0.959 &      0.211 &      0.000 &     11.064 \\

$\Gamma_8$ &      2.969 &      1.187 &      0.170 &      0.034 &     -1.069 & $\Gamma_8$ &      1.971 &      1.081 &      0.142 &      0.027 &      0.635 & $\Gamma_8$ &      1.891 &      1.160 &      0.157 &      0.209 &    -10.332 \\

$\Gamma_7$ &      3.775 &      0.182 &      0.157 &      0.003 &    -17.969 & $\Gamma_7$ &      3.153 &      0.180 &      0.141 &      0.004 &     76.350 & $\Gamma_7$ &      2.418 &      0.163 &      0.143 &      0.005 &     14.487 \\

$\Gamma_7$ &      4.822 &      0.924 &      2.729 &      0.017 &     26.128 & $\Gamma_7$ &      3.993 &      0.883 &      4.253 &      0.104 &      6.743 & $\Gamma_7$ &      3.384 &      1.155 &      1.549 &      0.054 &     16.784 \\

$\Gamma_9$ &      4.831 &      0.923 &      2.652 &      0.000 &     28.017 & $\Gamma_9$ &      4.133 &      0.857 &      5.281 &      0.000 &     31.568 & $\Gamma_9$ &      3.577 &      1.149 &      3.279 &      0.000 &     23.544 \\

$\Gamma_7$ &      5.193 &      2.263 &      0.409 &      0.018 &    -36.548 & $\Gamma_7$ &      4.360 &      2.806 &      0.484 &      0.219 &    -67.663 & $\Gamma_7$ &      3.872 &      3.705 &      0.551 &      0.210 &    -67.883 \\
\hline
\end{tabular}
\caption{Irreducible representations (IR) for zone center sates,
respective transition energies, effective masses,  linear and cubic
Dresselhaus coefficients terms for the WZ phase of AlP, AlAs and
AlSb. All transition energies are offset to the top of the valence
band. }
\label{tab:WZ-vals1}
\end{sidewaystable}
\begin{sidewaystable}
\begin{tabular}{||c|c|c|c|c|c||c|c|c|c|c|c||c|c|c|c|c|c||}
\hline \hline
\multicolumn{6}{||c|}{\bf GaP} & \multicolumn{6}{|c|}{\bf GaAs} & \multicolumn{6}{|c||}{\bf GaSb} \\
\hline
     $IR$ &  $E (eV)$ &  $m_{||}$ & $m_{\perp}$ & $\zeta_1 (eV\AA)$ & $\zeta_3 (eV\AA^3)$ &      $IR$ &  $E (eV)$ &  $m_{||}$ & $m_{\perp}$ & $\zeta_1 (eV\AA)$ & $\zeta_3 (eV\AA^3)$ &      $IR$ &  $E (eV)$ &  $m_{||}$ & $m_{\perp}$ & $\zeta_1 (eV\AA)$ & $\zeta_3 (eV\AA^3)$ \\
\hline
$\Gamma_7$ &   -12.942 &      1.391 &      1.400 &      0.000 &    -10.751 & $\Gamma_7$ &    -12.033 &      1.339 &      1.350 &      0.000 &    -28.285 & $\Gamma_7$ &    -10.757 &      1.270 &      1.288 &      0.000 &    -58.903 \\

$\Gamma_8$ &   -11.369 &      0.615 &      1.768 &      0.076 &     -0.070 & $\Gamma_8$ &    -10.452 &      0.533 &      1.520 &      0.003 &     -0.089 & $\Gamma_8$ &     -9.269 &      0.452 &      1.453 &      0.028 &     -0.301 \\

$\Gamma_8$ &    -6.313 &      0.312 &      2.362 &      0.064 &      0.003 & $\Gamma_8$ &     -6.289 &      0.291 &     12.730 &      0.188 &      0.467 & $\Gamma_8$ &     -5.992 &      0.261 &      3.827 &      0.344 &      1.577 \\

$\Gamma_8$ &    -1.327 &      1.577 &      0.242 &      0.105 &     18.694 & $\Gamma_8$ &     -1.291 &      1.698 &      0.242 &      0.440 &     31.168 & $\Gamma_8$ &     -1.537 &      1.535 &      0.193 &      1.137 &     25.843 \\

$\Gamma_9$ &    -1.233 &      1.624 &      0.238 &      0.000 &     21.321 & $\Gamma_9$ &     -1.049 &      1.745 &      0.224 &      0.000 &     42.790 & $\Gamma_9$ &     -1.030 &      1.604 &      0.131 &      0.000 &     77.808 \\

$\Gamma_7$ &    -0.373 &      0.118 &      1.145 &      0.068 &     26.433 & $\Gamma_7$ &     -0.475 &      0.118 &      0.434 &      0.048 &    141.143 & $\Gamma_7$ &     -0.874 &      0.149 &      0.436 &      0.147 &   -141.671 \\

$\Gamma_7$ &    -0.050 &      0.821 &      0.210 &      0.072 &     35.712 & $\Gamma_7$ &     -0.120 &      0.200 &      0.197 &      0.067 &     67.513 & $\Gamma_7$ &     -0.142 &      0.086 &      0.192 &      0.175 &     23.793 \\

$\Gamma_9$ &     0.000 &      0.941 &      0.205 &      0.000 &     29.339 & $\Gamma_9$ &      0.000 &      1.026 &      0.134 &      0.000 &     36.419 & $\Gamma_9$ &      0.000 &      0.833 &      0.087 &      0.000 &     73.448 \\

$\Gamma_8$ &     2.251 &      1.162 &      0.143 &      0.075 &     -3.725 & $\Gamma_8$ &      1.503 &      1.050 &      0.125 &      0.212 &    -12.750 & $\Gamma_8$ &      0.509 &      0.983 &      0.096 &      0.716 &    -72.842 \\

$\Gamma_7$ &     2.877 &      0.153 &      0.125 &      0.006 &     53.344 & $\Gamma_7$ &      1.588 &      0.090 &      0.082 &      0.037 &    -55.218 & $\Gamma_7$ &      0.851 &      0.064 &      0.060 &      0.034 &    -33.327 \\

$\Gamma_7$ &     4.395 &      1.135 &      1.665 &      0.076 &     15.714 & $\Gamma_7$ &      4.271 &      0.861 &      1.977 &      0.355 &     62.864 & $\Gamma_7$ &      2.824 &      0.857 &      1.225 &      0.184 &     71.733 \\

$\Gamma_9$ &     4.429 &      1.233 &      1.872 &      0.000 &      9.533 & $\Gamma_9$ &      4.417 &      0.793 &      0.722 &      0.000 &    123.206 & $\Gamma_9$ &      2.970 &      0.785 &      0.695 &      0.000 &     57.196 \\

$\Gamma_7$ &     4.940 &      1.510 &      0.608 &      0.004 &    -76.564 & $\Gamma_7$ &      4.575 &      0.974 &      0.317 &      0.337 &   -282.988 & $\Gamma_7$ &      3.284 &      0.394 &      0.470 &      0.262 &   -283.959 \\
\hline
\end{tabular}
\caption{Irreducible representations (IR) for zone center sates,
respective transition energies, effective masses,  linear and cubic
Dresselhaus coefficients terms for the WZ phase of GaP, GaAs and
GaSb. All transition energies are offset to the top of the valence
band. }
\label{tab:WZ-vals2}
\end{sidewaystable}
\begin{sidewaystable}
\begin{tabular}{||c|c|c|c|c|c||c|c|c|c|c|c||c|c|c|c|c|c||}
\hline \hline
\multicolumn{6}{||c|}{\bf InP} & \multicolumn{6}{|c|}{\bf InAs} & \multicolumn{6}{|c||}{\bf InSb} \\
\hline
     $IR$ &  $E (eV)$ &  $m_{||}$ & $m_{\perp}$ & $\zeta_1 (eV\AA)$ & $\zeta_3 (eV\AA^3)$ &      $IR$ &  $E (eV)$ &  $m_{||}$ & $m_{\perp}$ & $\zeta_1 (eV\AA)$ & $\zeta_3 (eV\AA^3)$ &      $IR$ &  $E (eV)$ &  $m_{||}$ & $m_{\perp}$ & $\zeta_1 (eV\AA)$ & $\zeta_3 (eV\AA^3)$ \\
\hline
$\Gamma_7$ &    -11.746 &      1.660 &      1.677 &      0.000 &    -28.484 & $\Gamma_7$ &    -11.875 &      2.071 &      2.034 &      0.001 &    -30.395 & $\Gamma_7$ &    -10.181 &      1.802 &      1.778 &      0.001 &    -40.757 \\
$\Gamma_8$ &    -10.711 &      0.936 &      2.509 &      0.036 &     -0.038 & $\Gamma_8$ &    -11.151 &      1.397 &      3.377 &      0.034 &     -0.116 & $\Gamma_8$ &     -9.421 &      1.094 &      2.932 &      0.043 &     -0.155 \\
$\Gamma_8$ &     -5.091 &      0.390 &      1.845 &      0.087 &      0.223 & $\Gamma_8$ &     -4.526 &      0.469 &      1.349 &      0.088 &      0.248 & $\Gamma_8$ &     -4.351 &      0.420 &      1.244 &      0.004 &     -0.317 \\
$\Gamma_8$ &     -0.949 &      1.833 &      0.237 &      0.011 &      3.746 & $\Gamma_8$ &     -0.910 &      2.116 &      0.210 &      0.078 &      5.879 & $\Gamma_8$ &     -1.021 &      2.185 &      0.222 &      0.188 &      2.114 \\
$\Gamma_9$ &     -0.849 &      1.894 &      0.230 &      0.000 &      2.693 & $\Gamma_9$ &     -0.652 &      2.164 &      0.166 &      0.000 &     27.849 & $\Gamma_7$ &     -0.847 &      0.203 &      0.210 &      0.595 &    633.586 \\
$\Gamma_7$ &     -0.348 &      0.097 &      1.205 &      0.074 &     97.187 & $\Gamma_7$ &     -0.469 &      0.115 &      0.319 &      0.849 &    744.548 & $\Gamma_9$ &     -0.508 &      2.310 &      0.206 &      0.000 &      7.028 \\
$\Gamma_7$ &     -0.063 &      0.839 &      0.169 &      0.084 &     65.030 & $\Gamma_7$ &     -0.105 &      0.101 &      0.113 &      1.414 &    648.740 & $\Gamma_7$ &     -0.098 &      0.058 &      0.094 &      1.812 &    178.644 \\
$\Gamma_9$ &      0.000 &      1.273 &      0.158 &      0.000 &     45.639 & $\Gamma_9$ &      0.000 &      1.700 &      0.084 &      0.000 &   1107.720 & $\Gamma_9$ &      0.000 &      2.060 &      0.066 &      0.000 &   2450.059 \\
$\Gamma_7$ &      1.474 &      0.105 &      0.088 &      0.011 &    -54.015 & $\Gamma_7$ &      0.481 &      0.060 &      0.042 &      0.571 &  -1143.621 & $\Gamma_7$ &      0.287 &      0.051 &      0.035 &      1.212 &  -2955.560 \\
$\Gamma_8$ &      1.712 &      1.094 &      0.132 &      0.032 &     -1.600 & $\Gamma_8$ &      1.222 &      1.276 &      0.113 &      0.007 &     -1.456 & $\Gamma_8$ &      1.116 &      1.781 &      0.118 &      0.197 &    -11.492 \\
$\Gamma_7$ &      4.535 &      1.646 &      0.952 &      0.097 &     70.102 & $\Gamma_7$ &      4.445 &      3.377 &      0.255 &      0.790 &   -106.736 & $\Gamma_7$ &      3.395 &      1.785 &      2.105 &      1.909 &    -44.165 \\
$\Gamma_9$ &      4.575 &      1.701 &      0.845 &      0.000 &     95.004 & $\Gamma_7$ &      4.631 &      2.025 &      0.059 &      0.791 &  -2927.150 & $\Gamma_7$ &      3.513 &      0.656 &      1.635 &      1.987 &     42.178 \\
$\Gamma_7$ &      4.802 &      0.804 &      0.448 &      0.158 &    -98.388 & $\Gamma_9$ &      4.662 &      2.580 &      0.052 &      0.000 &   2824.700 & $\Gamma_9$ &      3.930 &      0.565 &      2.502 &      0.000 &     38.729 \\
\hline
\end{tabular}
\caption{Irreducible representations (IR) for zone center sates,
respective transition energies, effective masses,  linear and cubic
Dresselhaus coefficients terms for the WZ phase of InP, InAs and
InSb. All transition energies are offset to the top of the valence
band. }
\label{tab:WZ-vals3}
\end{sidewaystable}
\begin{table}[h]
\begin{center}
\begin{tabular}{c|c|c|c|c}
\hline \hline
     & $E_g$ (eV) & $\Delta_{so}$ (eV) & $\Delta_{cr}$ (eV) & $\Delta{E}_{CB}$ (eV) \\
\hline
       AlP &      $2.969~(\Gamma_8)$ &      0.070 &      0.409 &    -0.1428 \\
      AlAs &      $1.971~(\Gamma_8)$ &      0.319 &      0.338 &    -0.0841 \\
      AlSb &      $1.891~(\Gamma_8)$ &      0.683 &      0.276 &    -0.0909 \\
       GaP &      $2.251~(\Gamma_8)$ &      0.082 &      0.341 &    -0.0801 \\
      GaAs &      $1.503~(\Gamma_8)$ &      0.351 &      0.244 &    -0.0632 \\
      GaSb &      $0.509~(\Gamma_8)$ &      0.777 &      0.239 &    -0.1186 \\
       InP &      $1.474~(\Gamma_7)$ &      0.108 &      0.303 &    -0.0646 \\
      InAs &      $0.481~(\Gamma_7)$ &      0.379 &      0.195 &     0.0405 \\
      InSb &      $0.287~(\Gamma_7)$ &      0.787 &      0.159 &     0.0872 \\
\hline
\end{tabular}
\caption{ Energies of the direct gap III-V
WZ semiconductors. The symmetry of the conduction band minimum is indicated with the band gap. $\Delta_{so}$ and $\Delta_{cr}$ are the spin-orbit splitting and crystal-field splitting energies extracted using Eq.~(\ref{WZ:CrSo}). $\Delta{E}_{VB}=E_{VB}^{ZB}-E_{VB}^{WZ}$, is the energy difference between the top of the valance bands for
the two polytypes.}
\label{tab:WZ-compact}
\end{center}
\end{table}
\begin{center}
\begin{table*}[ht]
\begin{tabular}{c|c|c|c}
\hline
     & $E_g$(eV) Present Calculations & $E_g$(eV) {\it Ab initio} Methods & $E_g$(eV) Experiment \\
\hline

      GaAs &      $1.503~(\Gamma_8)$ &      1.623[\cite{Yeh1994}, 1.6\cite{Murayama1994}, 1.381\cite{Zanolli2007prb}$^a$,1.811\cite{Zanolli2007prb}$^b$  & 1.498\cite{Haraguchi1994},1.476\cite{Soshnikov2006},1.494\cite{Regolin2007},1.51\cite{Khorenko2004}      \\

       InP &      $1.474~(\Gamma_7)$ &      1.5403\cite{Murayama1994} &      1.44\cite{Mattila2007},1.452\cite{Reitzenstein2007},1.49\cite{Mishra2007},1.508\cite{Ding2007},1.645\cite{Tragardh2007}      \\

      InAs &      $0.481~(\Gamma_7)$ &      0.6424\cite{Murayama1994}, 0.47\cite{Zanolli2007prb} &      0.54\cite{Tragardh2007}      \\
\hline
\end{tabular}
\caption{ A comparison between WZ band gaps from empirical pseudopotentials, those obtained from first principle calculations, and experimental results. The symmetry of the conduction band minimum is indicated with the band gap.
The {\it ab initio} results are LDA calculations from Refs. \onlinecite{Murayama1994} and \cite{Yeh1994},  and GW and SX results from Ref. \cite{Zanolli2007prb}. In all three materials our results are in agreement with the experimental values and trends. Note, that the  {\it ab initio} values listed above are not directly contained in the references, but are obtained from the calculated difference in the band gap between the two polytypes.
}
\label{tab:gap-compare}
\end{table*}
\end{center}

The spin-orbit coupling can alter the ordering of the valence band states in different WZ semiconductors.
For example, in CdS and CdSe  the top three valence states are in descending order $\Gamma_9$, $\Gamma_7$, $\Gamma_7$  (referred to as  normal ordering)\cite{Birman1959b,Thomas1959} while in ZnO the ordering is $\Gamma_7$, $\Gamma_9$, $\Gamma_7$ (anomalous ordering) which results from a negative spin-orbit energy. In our calculations, all materials except InSb  have normal ordering.
In the case of InSb the ordering of the valence band states is complicated by its very large spin-orbit splitting which forces the $\Gamma_7$ split-off hole bellow the next $\Gamma_9$ state (which comes from the 'folded over' $p$-like $L$-valley states in ZB).
This results in the unusual $\Gamma_9$, $\Gamma_7$, $\Gamma_9$, $\Gamma_7$ ordering of valence band states in InSb.

Tables \ref{tab:WZ-vals1}-\ref{tab:WZ-vals3} also give the Dresselhaus coefficients for each band.
As expected\cite{LewYanVoon1996}, all $\Gamma_9$ states have zero linear Dresselhaus coefficients ($\zeta_1$) while all $\Gamma_7$  and $\Gamma_8$ states have nonzero linear and cubic spin-splitting coefficients.
Amongst  the five WZ phase semiconductors with the largest spin-orbit energies ($\Delta_{so}$), InSb has the largest Dresselhaus coefficients, followed by GaSb, InAs, GaAs and AlSb.

\section{Summary}\label{sec:summary}
We have calculated the electronic band structure for nine III-V semiconductors in WZ phase using empirical pseudopotentials
with the inclusion of spin-orbit coupling. The predicted band
structures are based on the concept of transferable model
potentials. Our calculations show that in WZ phase, InP, InAs and
InSb have a direct gap ($\Gamma_7$) which is larger than the corresponding zincblende material.
AlP, AlAs, AlSb, GaP, GaAs and GaSb also have direct gaps, but with $\Gamma_8$ conduction band minima.
WZ AlP and AlAs have  larger direct gaps than the indirect gaps of their ZB polytypes.
 The opposite trend is seen in AlSb and GaP which have smaller direct gaps than their indirect gaps in ZB phase.
In WZ phase GaAs and GaSb have direct gaps which are smaller than their ZB counterparts.

Our calculations are in excellent agreement with with experimentally obtained band gaps for GaAs, InAs
and InP in WZ phase.
Significantly, our results agree with experiment over whether the WZ band gap is larger or smaller than the ZB gap, in contrast to {\it ab initio} methods.
We have extracted the linear and cubic Dresselhaus spin-splitting
coefficients
and find they are generally, though not always, larger for materials with larger spin-orbit coupling.
The relatively large spin-splittings may be of use for spin-dependent transport in WZ nanowires.
More recently, the WZ phase of
GaAs has been grown in bulk\cite{McMahon2005}. More experimental
measurements on such bulk WZ phase III-V semiconductors, would lead
to a clearer understanding of their electronic properties.

\section{Acknowledgements}\label{sec:summary}

We would like to acknowledge support from the University of Iowa.


\end{document}